\documentclass[twocolumn,tighten]{aastex63}
\usepackage{amsmath}
\usepackage{graphicx}
\newcommand\e[1]{\times 10^{#1}}
\newcommand\MeijerG[7]{G_{#3,#4}^{#1,#2} \left( #7 \left|
\begin{array} {c}#5 \\ #6 \\ \end{array} \right.\right)}
\newcommand\MeijerGFour[9]{G_{#3,#4}^{#1,#2} \left( #9 \left|
\begin{array} {cc}#5 & #6 \\ #7 & #8\\ \end{array} \right.\right)}
\newcommand\GMeijerG{\MeijerG{m}{n}{p}{q}{a_1, \dots, a_p}{b_1, \dots, b_q}{z}}
\usepackage{xcolor}

\begin{document}
\title{Testing Emergent Gravity with Isolated Dwarf Galaxies}
\author{Kris Pardo}\email{kpardo@caltech.edu}
\affiliation{Jet Propulsion Laboratory, California Institute of Technology\\ Pasadena, CA 91101, USA}
\affil{Department of Astrophysical Sciences, Princeton University \\ 4 Ivy Lane, Princeton, NJ 08544}

\begin{abstract}
Verlinde (2016) has proposed a new modified theory of gravity, Emergent Gravity (EG), as an alternative to dark matter. EG reproduces the Tully-Fisher relationship with no free parameters and agrees with the velocity curves of most massive, spiral galaxies well. In its current form, the theory only applies to isolated, spherically symmetric systems in a dark energy-dominated Universe, and thus can only be tested fairly with such systems. This paper presents a framework for rotation curve tests of EG using isolated dwarf galaxies. Here I extend the EG equations to axisymmetric distributions for the first time. I also perform a preliminary test of the predictions from EG versus the maximum velocity measurements of 452 isolated dwarf galaxies. I find that EG predicts the maximum velocities of these systems somewhat well for galaxies with measured velocities around $100~\rm{km/s}$. EG severely underpredicts the maximum velocities for dwarf galaxies with measured velocities greater than this value and overpredicts for those with measured velocities less than this value. Rotation curves of these isolated dwarf galaxies would provide the definitive test of EG.
\vspace{1cm}
\end{abstract}

\section{Introduction}
Since Zwicky's paper on ``dunkel matter'' in the 1930s \citep{Zwicky}, there has been growing evidence for new physics. Rotation curves of galaxies \citep{Rubin1980}, the Cosmic Microwave Background \citep{Planck2015}, and Baryonic Acoustic Oscillations \citep{Eisenstein2005} have solidified the need for something beyond the Standard Model or General Relativity (GR). Is this new physics a dark matter (DM) particle or a modification of GR? \cite{Verlinde2016} gives one possible solution to this question.

In Emergent Gravity (EG) theory \citep{Verlinde2016}, gravity emerges from the entanglement of spacetime. According to this theory, dark energy has some entanglement entropy. Baryonic matter displaces dark energy and, due to the volume law contribution to entropy, this causes an elastic response force on the matter. This manifests itself as an extra gravitational force around massive objects. \cite{Verlinde2016} uses this elastic response force ansatz to produce an equation for the ``apparent dark matter'' given some baryonic mass distribution. 

In the limit of a point-source mass, the equation for the apparent DM in EG converges to the weak limit equation from Modified Newtonian Gravity (MOND) \citep{Milgrom1983}. Thus, \cite{Verlinde2016} manages to derive the Tully-Fisher relation within his theory using \emph{no free parameters} and directly connects the MOND acceleration, $a_0$, to the energy density in dark energy.

However, EG in its current formulation only applies to the current, deSitter-like Universe. The equations given in \cite{Verlinde2016} are also only valid for spherically symmetric, isolated systems. Nonetheless, there have been several tests of this theory. \cite{Brouwer2016} studied the weak lensing of galaxy clusters, and found it to be consistent with EG. \cite{Ettori2016} found EG to agree with two large, roughly spherical galaxy clusters, and \cite{Diez2016} also found agreement with the mass-to-light ratios of the classical dwarf spheroidal satellite galaxies. Several studies claim that EG is inconsistent with observations: the initial mass functions of massive early-type galaxies \citep{Tortora2017}, the radial acceleration within the inner regions of spiral galaxies \cite{Lelli2017}, and the perihelia of Solar system planets \cite{Hees2017}. However, all of these tests attempt to apply EG outside of the currently narrow regime where it makes robust predictions: spherically symmetric, isolated systems in the nearby Universe. 

In this paper, I extend the EG formalism to axisymmetric systems and consider its predictions for the maximum velocities within isolated dwarf galaxies. I then compare these predictions to observations. These systems fulfill all of the requirements of the current formulation of EG, and thus provide the strongest constraints on EG. In Section 2, I derive the equations for specific spherical and axisymmetric mass distributions along with the corresponding ``apparent'' dark matter predicted from EG. In Section 3, I describe how I apply these equations to isolated dwarf galaxies. In Section 4, I compare EG's predictions for the velocities within isolated dwarf galaxies to those measured in a recent 21 cm study \citep{Bradford2015}. I discuss these results and conclude in Section 5.

\section{Apparent DM Distribution Predictions from EG for Two Realistic Baryonic Mass Distributions}\label{sec:theory}
The goal of this section is to describe the velocity curve for two extended mass distributions in EG. Conservation of energy tells us that the circular velocity, $v(r)$, is given solely by the mass distribution. For standard $\Lambda$CDM, we simply use the mass distributions of both the baryonic and DM mass. In EG, we instead derive the apparent dark matter mass distribution from the baryonic mass distribution and then use both of these to find the velocity curve. In this section, I will derive the apparent dark matter mass distributions for both a spherically symmetric baryonic mass distribution and a axisymmetric baryonic mass distribution.

\subsection{Spherically Symmetric Case}\label{sec-sphere}

For a spherically symmetric, isolated system, the apparent DM predicted by EG \citep{Verlinde2016} is
\begin{equation}\label{eqn:verlinde}
\int_0^r \frac{GM_D^2 (r')}{r'^2} dr' = \frac{a_0 r}{6} M_B(r) \; ,
\end{equation}
where $G$ is Newton's gravitational constant, and $a_0 = c H_0$. By taking the derivative of both sides with respect to $r$, we find an equation for $M_D(r)$,
\begin{equation}\label{eqn:md}
M_D^2 (r) = \frac{a_0 r^2}{6G} \frac{d}{dr} \Big( r M_B(r) \Big) \; .
\end{equation}
Note that if we allow $M_B$ to be a point-mass, then $M_D^2(r) = \frac{a_0 r^2}{6G}M_B$, which would give a gravitational acceleration of
\begin{equation}\label{eqn:mondg}
g_D(r) = \frac{GM_D(r)}{r^2} = \sqrt{\frac{a_0}{6} g_B(r)} \; .
\end{equation}
This is just the MOND acceleration in the weak-field limit \citep{Milgrom1983} with $a_M = \frac{a_0}{6}$. I only include this as an aside -- dwarf galaxies are of course not describable as point-masses.

Instead, let us consider an extended mass distribution. In particular, let us employ a deprojected S\'{e}rsic profile. These profiles fit the stellar light of galaxies well, and since we are assuming there is no dark matter, this should also be a good measure of the mass.

The S\'{e}rsic profile of a galaxy is given by
\begin{equation}\label{eqn:sersic}
I(R) = I_e \exp \left[ 1-b_n \left(\frac{R}{R_e}\right)^{1/n} \right] \; ,
\end{equation}
where $I_e$ and $R_e$ are the intensity and projected radius at the half-light slice, respectively, and $n$ is the so-called S\'{e}rsic index, which is a measure of the concentration of the light about the center. The constant $b_n$ is given by gamma functions (see Appendix \ref{app:sersic}). 

To find the mass profile, we must first deproject the S\'{e}rsic profile to give the luminosity density. Assuming spherical symmetry, we can then integrate in the angular directions to give the radial luminosity profile. \cite{MazureCapelato2002} first found the exact solution for the radial luminosity profile given a general S\'{e}rsic profile, and I use their results here.

Since we are assuming that there is no dark matter, the mass must follow the light. Then, the stellar mass profile should be the same as the luminosity profile except for some scaling factor, the baryonic mass-to-light ratio, $\Upsilon$. This ratio, along with the effective intensity simply give the normalization of the function, and thus we let $\Sigma = I_e\Upsilon$, where the process for setting this normalization constant is given in Section \ref{sec:isodata}. The final equation for the baryonic mass profile is
\begin{equation}\label{eqn:sersicmr}
\begin{aligned}
M_B(r) = & 2\pi c_1 \Sigma R_e^2 \left(\frac{r}{R_e}\right)^{\frac{2n+1}{n}} \\
\times & \MeijerGFour{2n}{1}{1}{2n+1}{\{-\left(\frac{1}{2n}\right)\},}{\{\}}{\{\beta_s\},}{\{-\left(\frac{2n+1}{2n}\right)\}}{c_2\left(\frac{r}{R_e}\right)^2} \; ,
\end{aligned}
\end{equation}
where $\GMeijerG$ is the Meijer G function (described in Appendix \ref{sec:meijerg}), and the $c_1$, $c_2$, and $\beta_s$ are constants (described in Appendix \ref{app:radialmassprof}).
Then, the apparent DM predicted by EG due to this realistic mass distribution is given exactly by
\begin{equation} \label{eqn:sersicdm}
\begin{aligned}
M_{D}^2(r) = & \frac{\pi a_0 c_1 \Sigma R_e^2}{3G}\left(\frac{r}{R_e}\right)^{\frac{2n+1}{n}} r^2 \\
\times & \Bigg[\MeijerGFour{2n}{1}{1}{2n+1}{\{-\left(\frac{1}{2n}\right)\},}{\{\}}{\{\beta_s\},}{\{-\left(\frac{2n+1}{2n}\right)\}}{c_2 \left(\frac{r}{R_e}\right)^2} \\
+ \ & 2 \MeijerGFour{2n}{0}{0}{2n}{\{\}}{\{\}}{\{\beta_s\}}{\{\}}{c_2 \left(\frac{r}{R_e}\right)^2} \Bigg] \; .
\end{aligned}
\end{equation}
For a detailed description of these methods, see Appendix \ref{app:sersic}.

Since this is a spherically symmetric mass distribution, the circular velocity is given by
\begin{equation}\label{eqn:vreg}
v(r) = \pm \sqrt{\frac{G(M_{D}(r) + M_{B}(r))}{r}} 
\end{equation}

\subsection{Axisymmetric Case}\label{sec-disk}
Now let us consider the axisymmetric case. The EG equation for axisymmetric systems was not given in \cite{Verlinde2016}; however, it is straightforward to derive from the general equations included in the article. \cite{Verlinde2016} gives the general form of the EG equations in 3+1 dimensions as
\begin{equation}\label{eqn:verlindeequation}
\int_B \left( \frac{8\pi G}{a_0} \Sigma_D\right)^2 dV = \frac{2}{3} \int_{\partial B} \frac{\Phi_B}{a_0} n_i dA_i \; ,
\end{equation}
where $B$ is an arbitrary integration region, $n_i$ is a vector normal to the surface of the region, $\Sigma_D$ is the mass surface density of the apparent dark matter, $\Phi_B$ is the gravitational potential induced by the baryonic matter, and $a_0 = c/H_0$ is the acceleration constant, as before.

Without specifying a mass distribution, other than assuming it is integrable, we can rewrite this as
\begin{equation}\label{eqn:geneg} 
\Sigma_D^2 = \frac{a_0}{96 \pi^2 G^2} \left[ \nabla \cdot (\Phi_B n_i)\right] \; .
\end{equation}
We can now allow for any baryonic mass distribution and solve for the apparent dark matter mass distribution. To keep the problem tractable, I will use an axisymmetric distribution -- specifically, an exponential mass distribution. The surface density for an exponential mass distribution is given by
\begin{equation}
\Sigma(R) = \Sigma_d e^{-R/R_d} \; ,
\end{equation}
where $R_d$ is the scale length of the disk and $\Sigma_d$ is a normalization constant. Note that this distribution is often used to describe the mass in disk systems.

It is straightforward to find the potential for this system via a Hankel Transform\cite[e.g.][]{BT}. It is given as
\begin{equation}
\Phi = \Phi(R,z) = -2\pi G \Sigma_d R_d^2 \int_0^{\infty} J_0(kR)\frac{\exp(-k|z|)}{(1+k^2R_d^2)^{3/2}} \; ,
\end{equation}
where $J_n(kR)$ is a bessel function.
For the circular velocity, we only need to concern ourselves with the mass distribution in the plane. Thus, we will only consider the baryonic potential in the plane. This leads to an apparent DM mass surface density of
\begin{multline}
\Sigma_D^2(R,0) = \frac{a_0}{96\pi G} \frac{\Sigma_d}{R_d}\Big[ I_0\left(y \right)\left(R_dK_1\left(y\right) - R K_0\left(y \right)\right)\\ - I_1\left(y \right) \left(R_d K_0\left(y\right) - R K_1\left(y\right)\right)\Big] \; ,
\end{multline}
where $I_n(x)$ and $K_n(x)$ are modified bessel functions and $y=R/2R_d$.
Finally, we find our circular velocity through the equation
\begin{equation}
v_c(R) = R \left. \frac{\partial (\Phi_T)}{\partial R}\right|_{z=0} \; ,
\end{equation}
where $\Phi_T$ is the total potential. We find this by adding all sources of the surface density $\Sigma_T = \Sigma_B + \Sigma_D$ and then transforming to the potential using a Hankel Transform.

\section{Modeling Isolated Dwarf Galaxy Rotation Curves with EG}
In this section, I apply my equations from Section \ref{sec:theory} to real isolated dwarf galaxies. First, I describe the equations employed in the analysis and then I discuss the data.

\subsection{Theory}
Isolated dwarf galaxies contain a significant amount of HI gas that often exceeds the amount of stellar mass in the galaxy \citep{Geha2006}. This HI gas in dwarf galaxies typically extends far beyond the stellar disk \citep{Broeils1997}. Thus, we must include the mass profiles of both the stellar mass and the HI gas mass to properly model the baryonic content of these galaxies.

In addition, real galaxies are neither perfect spheres nor infinitely thin disks. These two cases form bounding cases for the possible velocities EG predicts. The true distribution is most likely something in between these two cases. Thus, I will develop the equations for both of these cases.

For the spherical case, I model the starlight profile as a S\'{e}rsic profile with index, $n$. I model the HI mass profile as a sphere with an exponential density profile -- a S\'{e}rsic profile with $n=1$. The scale lengths for each case, $R_{\star}$ and $R_{\rm{HI}}$, and the normalization constants, $\Sigma_{\star}$ and $\Sigma_{\rm{HI}}$, are given by measured quantities, as described in the next section.

In the axisymmetric case, I use an exponential surface density distribution for both the stars and the gas. The main parameters in each case, $R_{\star}$, $R_{\rm{HI}}$, $\Sigma_{d, \star}$, and $\Sigma_{d, \rm{HI}}$, are also derived from measured quantities. Note that the effective radii used in the spherical and axisymmetric cases are the same for each species, but the normalization constants are different. 

\subsection{Data}\label{sec:isodata}
To test EG, I use the \cite{Bradford2015} sample of isolated dwarf galaxies in SDSS DR 8. They choose all galaxies within the NASA Sloan Atlas\footnote{http://www.nsatlas.org} (NSA) catalog \citep{Blanton2011} that have $z>0.002$ and $M_r <17.72$. They then select according to an isolation criteria: for stellar mass $M_{\star} < 10^{9.5} M_{\odot}$, a galaxy is isolated if $d_{\rm{host}} > 1.5 \ \rm{Mpc}$. The full \cite{Bradford2015} sample has $546$ isolated dwarf galaxies ($M_{\star}<10^{9.5} \ M_{\odot}$). For each of these galaxies, \cite{Bradford2015} measure the 21 cm peak flux and line width. The HI gas masses are calculated from the peak fluxes. The inferred maximum circular velocity in each galaxy is given by
\begin{equation}
    v_{\rm{max}} = \frac{W_{20}}{2\sin i\ (1+z)} \; ,
\end{equation}
where $W_{20}$ is the width of the 21 cm line at $20\%$ peak flux, $i$ is the inclination of the galaxy, and $z$ is the redshift.

It has been found that face-on galaxies (i.e. galaxies with inclinations below $\sim 40$ degrees), can have significant errors induced by inclination effects \citep[c.f.][]{Stark2009}. To mitigate any effects from inclination, I select all galaxies with inclinations $i > 45$ degrees from the \cite{Bradford2015} sample. This leaves us with a final sample of $452$ galaxies.

For each of the galaxies in the sample, I use the NSA catalog S\'{e}rsic fit values for $n$, $R_{\star}$, and $M_{\star}$, and I use the \cite{Bradford2015} values for the HI mass and the measured maximum circular velocities. There are no direct observations of the normalization constants, $\Sigma_{\star}$, $\Sigma_{d,\star}$, $\Sigma_{\rm{HI}}$, and $\Sigma_{d, \rm{HI}}$. Instead these must be inferred from other quantities. I set each normalization constant by assuming that the measured mass is contained within five effective radii. Both of the distributions I use in this paper converge to their total masses with these radii. The normalization constants are different for the spherical and axisymmetric cases because of the different mass formulas assumed in each case.

To set the effective radius of the HI gas, $R_{\rm{HI}}$, I employ the relation by \cite{Lelli2016}:
\begin{equation}
\log_{10} M_{\rm{HI}} = (1.87\pm 0.03) \log_{10} R_{\rm{HI}} - (7.20 \pm 0.03) \; ,
\end{equation}
where $M_{\rm{HI}}$ is given in solar masses, and $R_{\rm{HI}}$ is given in kpc. The intrinsic scatter of the relation is $\sigma_{\rm{int}} = 0.06 \pm 0.01$ dex.

\section{Results}\label{sec:results}
Here I present the velocity curves predicted by EG. I also compare the predicted maximum circular velocities from EG to those measured in \cite{Bradford2015}. This is a preliminary analysis and should be followed by a full analysis with rotation curves of these galaxies.

To give an idea of the typical velocity curve produced by EG, let us consider the velocity curve of a sample dwarf galaxy with the median values from the data described in Section \ref{sec:isodata}. These values are all given in  Table \ref{tab-medianvalues}.

\begin{deluxetable}{cc}
\tablecolumns{2}
\tablewidth{\textwidth}
\tablecaption{Median Values for Isolated Dwarf Galaxies in Sample \label{tab-medianvalues}}
\tablehead{
\colhead{Parameters} &
\colhead{Values}}
\startdata
$M_{\star}$ & $3.98\e{8}\ M_{\odot}$\\
$R_{\star}$ & $2.07\ \rm{kpc}$\\
$n$ & $1.14$\\
$\Sigma_{\star}$ & $1.55\e{7}\ M_{\odot}\ \rm{kpc}^{-2}$\\
$M_{\rm{HI}}$ & $1.24\e{9}\ M_{\odot}$\\
$R_{\rm{HI}}$ & $10.31\ \rm{kpc}$\\
$\Sigma_{\rm{HI}}$ & $1.94\e{6}\ M_{\odot}\ \rm{kpc}^{-2}$\\
$v_{\rm{meas}}$ & $82\ \rm{km/s}$
\enddata
\end{deluxetable}

\begin{figure}[h]
\centering
\includegraphics[width=0.5\textwidth]{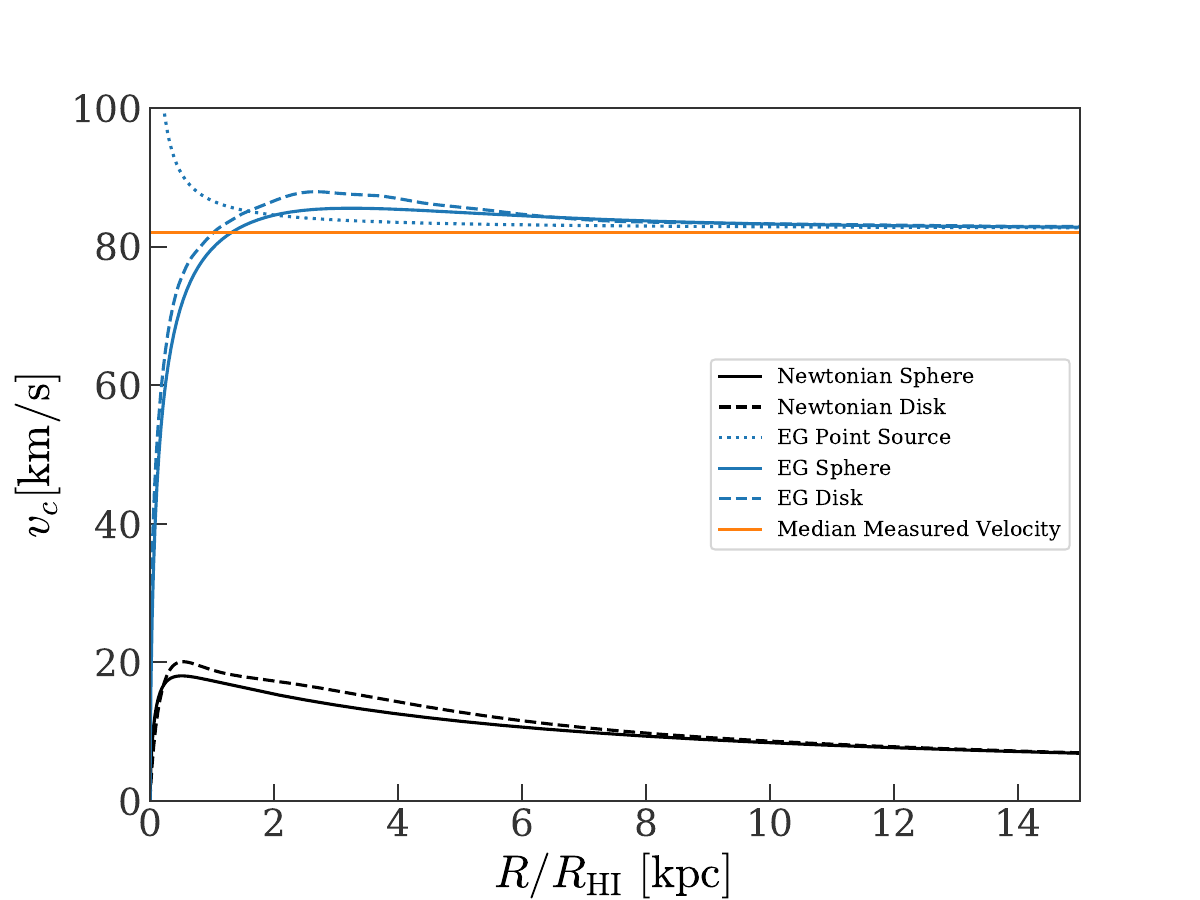}
\caption{\label{fig:vrth} Circular Velocity as a Function of Radius for a Median Isolated Dwarf Galaxy. The blue lines give the prediction from EG assuming a spherical baryonic mass distribution (solid line), an axisymmetric mass distribution (dashed line), or a point-mass (dotted line). The black lines give the Newtonian predictions (i.e. assuming there is only baryonic mass) for a spherical mass distribution (solid line) and an axisymmetric mass distribution (dashed line). The orange, dashed line gives the median measured maximum velocity from \cite{Bradford2015}.}
\end{figure}

The predicted velocity curves from EG for the spherical (solid), axisymmetric (dashed), and point-mass (dotted) cases are given as the blue curves in Figure \ref{fig:vrth}. For comparison, I also include the predictions from Newtonian gravity (assuming only baryonic matter), which are given by the black lines. The median measured maximum velocity from \cite{Bradford2015} is given by the orange, solid line as a reference. EG overpredicts in this median case.

Note that the maximum velocity for all of the cases occurs at $r\sim 5-30\ \rm{kpc}$. This is many times the effective radius of the stellar content. However, it is $\sim 0.5-3 R_{\rm{HI}}$. Thus, it is clear that the HI gas is the main driver behind the shape of the velocity curves, which agrees with the large gas fractions that are observed in these galaxies. 

\begin{figure}
\centering
\includegraphics[width=0.5\textwidth]{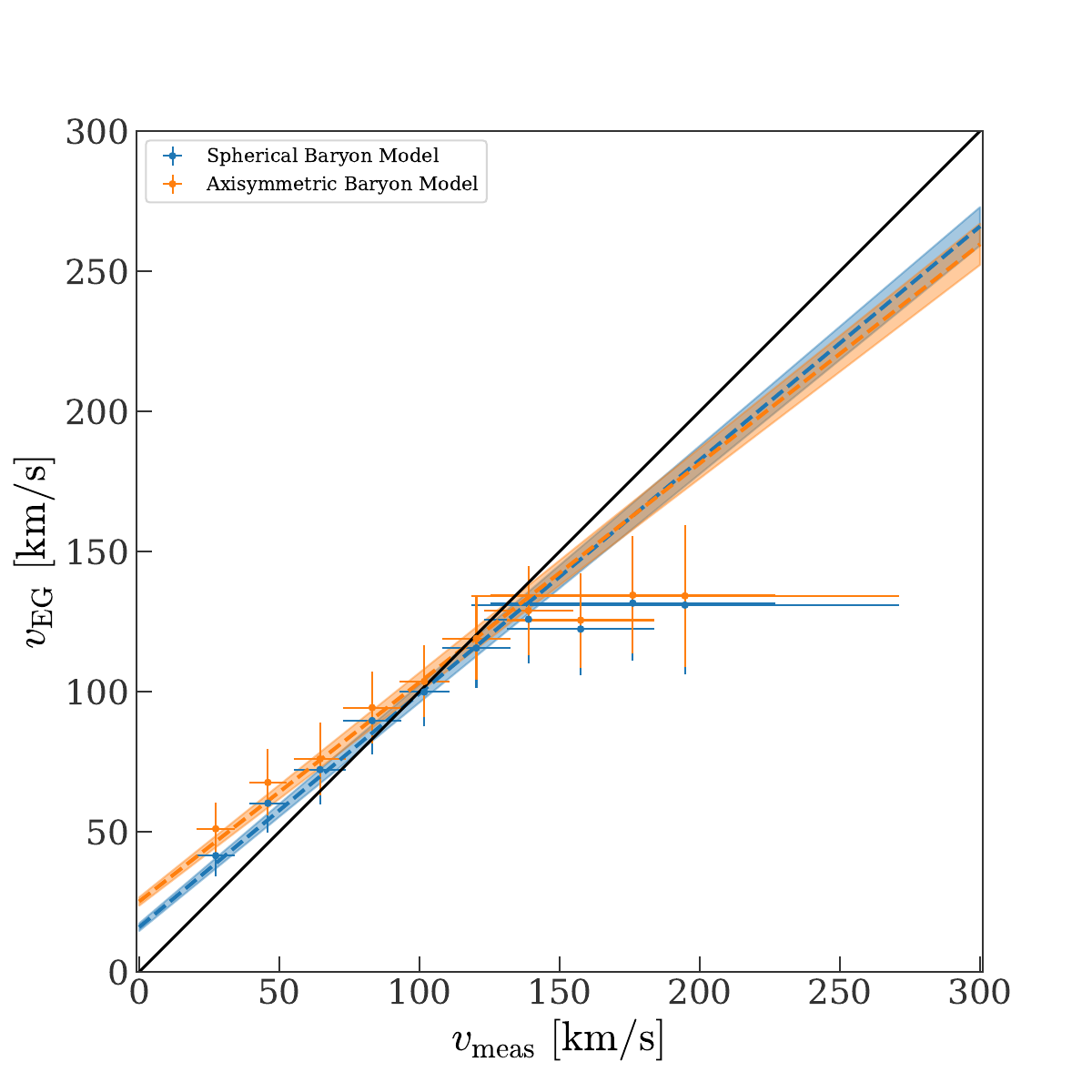}
\caption{\label{fig:isodwarfs} Binned, Predicted Maximum Circular Velocity from EG versus Measured Maximum Circular Velocities for Isolated Dwarf Galaxies. The blue points give the results for the spherically symmetric case and the orange points give the results for the axisymmetric case. If there was a perfect agreement with the measurements, all of the points would lie along the black line.}
\end{figure}

Figure \ref{fig:isodwarfs}, shows the binned, estimated maximum circular velocity from EG for the spherical case (blue) and axisymmetric case (orange) versus the measured maximum circular velocities from \cite{Bradford2015}. If the theory and observations were perfect, then all of the points would lie on the line $y=x$ (black line). While both models seem to do well around $100~\rm{km/s}$, they both overpredict the velocities at low measured $v$ and underpredict at high measured $v$.

I fit a best fit line to each of the models (assuming all of the galaxies can be treated independently) using a Markov chain Monte Carlo (MCMC) routine and plot these as the dashed lines in Figure \ref{fig:isodwarfs}. The best fit slope for the spherical model (blue) is $m=0.83\pm 0.03$ and the best fit intercept is $b=15.96\pm1.36$. The best fit values for the disk case (orange) are $m=0.78\pm 0.01$ and $b=25.12\pm1.51$. Neither of these models allow for the ``perfect agreement" line with $m=1$ and $b=0$. In fact, the disk case gives a slightly worse agreement with the data than the spherical case. Overall, these are preliminary results and a more careful analysis with rotation curves is needed to provide robust statements.

\section{Discussion \& Conclusions}
In this paper, I develop the equations for EG's velocity curve predictions based on two realistic baryonic mass profiles: a spherically symmetric model and an axisymmetric model. I then apply these models to isolated dwarf galaxies. These galaxies contain large amounts of HI gas, which must be treated separately from the stellar mass. Finally, I compare EG's predictions for the velocities with the HI 21-cm line width measurements from \cite{Bradford2015} for 452 isolated dwarf galaxies.

As I show in the results section, the predicted velocities from EG agree somewhat well for some ranges in measured $v$, but the predictions are systematically poor outside of this middle range. In particular, the very large measured velocities with $v>150~\rm{km/s}$ are not at all predicted by EG. Most of these galaxies are disks, so we would expect the axisymmetric models to better model the baryonic distribution of these galaxies. Again, these are preliminary results; however, it is intriguing that EG's predictions do not improve given improved modeling of the baryons.

There are many assumptions made when modeling the baryonic gas mass and it is entirely possible that any of these could be biasing the EG predictions. Perhaps the most error-prone parts of the analysis are the choices for the distributions and the normalization routine. 

The baryonic distributions I choose, a spherical distribution based on the S\'{e}rsic profile and an axisymmetric, infinitely-thin disk with an exponential distribution, are both highly idealistic. Nonetheless, they are the two extreme ends of the expected baryonic distributions for isolated dwarf galaxies -- thick disks. Since these galaxies are mostly composed of HI gas, we do expect them to be closer to thin disks than spheres. Specifically, HI gas tends to follow either an exponential or Gaussian distribution within galaxies \citep{Swaters2002, Martinsson2016}. Equation \ref{eqn:geneg} shows that the apparent dark matter surface density increases for larger baryonic potential flux losses through a surface. In other words, the steeper the baryonic distribution, the more apparent dark matter should be in the system, according to EG. Thus, using an exponential distribution rather than a Gaussian distribution for the disk case already gives an upper limit on the velocities EG would predict for an infinitely thin, axisymmetric disk. The use of a steeper distribution may alleviate some of the tension seen in Figure \ref{fig:isodwarfs}; however, it is not well supported by observational data.

As I describe in Section \ref{sec:isodata}, I normalize the mass distribution functions by assuming that all of the mass is contained within five effective radii. This normalization routine for each of the profiles is somewhat arbitrary, and it does have a large effect on the final predicted rotation curves. However, it is not clear that changing the normalization routine would allow the axisymmetric predictions to match the data while remaining consistent with observations of the HI content in disk galaxies. The velocities scale as $v \propto (\Sigma_{\rm{d},\star}+\Sigma_{\rm{d,\ HI}})^{1/2}$. To achieve the measured velocities, we would need to decrease the surface densities for lower-mass disks and increase the surface density for higher-mass disks. However, this does not agree with observations of HI in dwarf galaxies -- many surveys have found that there exists a tight relationship between the HI mass and effective radius \citep{Broeils1997, Lelli2016}. This implies a constant HI surface density. To be consistent with these observations, a constant normalization increase would need to be applied. This would only shift the points upward in Figure \ref{fig:isodwarfs} but would not change the shape of the relationship. Thus, either the lowest mass galaxies would have overpredicted velocities or the highest mass galaxies would have underpredicted velocities. 

In conclusion, I find hints of a discrepancy between the predicted maximum circular velocities from EG and the measured maximum circular velocities around isolated dwarf galaxies for the most realistic mass distributions. We need rotation curves of these galaxies to identify if this discrepancy is due to modeling errors or the inability of EG to describe these systems. Given that EG is only equipped to handle systems of this type, it seems that these discrepancies should be taken seriously as a possible issue with the theory. The next step is to obtain rotation curves of these isolated dwarf galaxies. The framework provided in this paper should allow for a robust test of EG with these rotation curves. This would provide the best test of EG at this time.

\acknowledgements
The author would like to thank David Spergel for his help and advice on this project. The author would also like to thank Jeremy Bradford, Erik Verlinde, Emmanuel Schaan, Adrian Price-Whelan, Andy Goulding, and Michael Strauss for helpful discussions. The author acknowledges support from the National Science Foundation Graduate Research Fellowship Program under grant DGE-1656466. This work was done as a private venture and not in the author's capacity as an employee of the Jet Propulsion Laboratory, California Institute of Technology.
\vfill

\newpage

\bibliographystyle{alpha}

\clearpage
\appendix

\section{Derivations of the EG Equations for a Deprojected S\'{e}rsic Profile} \label{app:sersic}
\subsection{The S\'{e}rsic Profile}

First, I repeat the equation describing the S\'{e}rsic profile
\begin{equation}\label{appe:sersic}
I(R) = I_e \exp \left[ 1-b_n \left(\frac{R}{R_e}\right)^{1/n} \right] \; , 
\end{equation}
where $b_n$ is defined by $\Gamma(2n) = 2\gamma(2n,b_n)$.

Note that $I(R)$ and $R$ are \textit{projected} quantities. They do not give the 3D, physical radius or intensity. To find the physical luminosity (and then the physical mass), we must deproject the S\'{e}rsic profile.

I begin by relating the intensity, $I(R)$, to the luminosity density, $n(r)$,
\begin{equation}
I(R) = 2 \int_0^{\infty} dz\ n(r) \; ,
\end{equation}
where I assume the luminosity density is symmetric in $z$. Note that $r$ is the radius in spherical coordinates and $R$ is the projected radius (i.e. the radius in cylindrical coordinates). Now, we can change variables using $r^2 = R^2 + z^2$. This gives
\begin{equation} \label{appe:intensityint}
I(R) = 2 \int_R^{\infty} \frac{r n(r)}{\sqrt{r^2-R^2}} \ dr \; .
\end{equation}
I find $n(r)$ by inverting Equation \ref{appe:intensityint} using the Abel Identity \citep[cf. Appendix B.5 of][]{BT}
\begin{equation}\label{appe:lumdensity}
n(r) = -\frac{1}{\pi} \int_r^{\infty} \frac{dI}{dR} \frac{dR}{\sqrt{R^2-r^2}}
\end{equation}
This is unsolvable for generic $I(R)$. However, the analytic solution to this integral for the $I(R)$ given in Equation \ref{appe:sersic} can be expressed in terms of Meijer G functions \citep{MazureCapelato2002}.

\subsection{The Meijer G functions}\label{sec:meijerg}
The Meijer G functions (see \url{http://functions.wolfram.com/HypergeometricFunctions/MeijerG/} and \url{http://dlmf.nist.gov/16} for more formulae involving the Meijer G functions) are generalized hypergeometric functions that give most of the special functions we know (i.e. trigonometric functions, Bessel functions, exponential function, etc.) as special cases. The Standard Meijer G function is defined as
\begin{equation}\label{appe:meijergdef}
\GMeijerG = \frac{1}{2\pi i} \int_L \frac{(\prod_{k=1}^m \Gamma(s+b_k))\prod_{k=1}^n \Gamma(1-a_k-s)}{(\prod_{k=n+1}^{p}\Gamma(s+a_k))\prod_{k=m+1}^q \Gamma(1-b_k-s)} z^{-s} ds
\end{equation}
A few useful identities of the Meijer G functions are [from DLMF]
\begin{eqnarray}\label{appe:mgidentity}
\GMeijerG &\equiv& z^{-c} \MeijerG{m}{n}{p}{q}{a_1+c, \dots, a_p+c}{b_1+c, \dots, b_q+c}{z}\\
\MeijerG{m}{n}{p}{q}{a_1, \dots, a_p}{b_1, \dots, b_q}{\frac{1}{z}} &\equiv& \MeijerG{n}{m}{q}{p}{1-b_1, \dots, 1-b_q}{1-a_1, \dots, 1-a_p}{z} \\
\GMeijerG &\equiv& \MeijerG{m}{n+1}{p+1}{q+1}{a_0,a_1, \dots, a_p}{b_1, \dots,b_q,a_0}{z}
\end{eqnarray}
The derivative of the Meijer G function leads to another Meijer G function [from Wolfram Functions]
\begin{equation}\label{appe:mgbasicderiv}
\frac{\partial \GMeijerG}{\partial z} = \MeijerG{m}{n+1}{p+1}{q+1}{-1,a_1-1, \dots, a_{n}-1, a_{n+1}-1,\dots, a_p-1}{b_1-1, b_m-1,0,b_{m+1}-1,\dots,b_q-1}{z}
\end{equation}
By combining Equations \ref{appe:mgbasicderiv} \& \ref{appe:mgidentity}, I find the following useful formula
\begin{equation}\label{appe:mgderiv}
\frac{\partial \left(z^{1-a_1} \GMeijerG \right)}{\partial z} = z^{-a_1}\MeijerG{m}{n}{p}{q}{a_1-1,a_2, \dots, a_p}{b_1, \dots, b_q}{z} \; .
\end{equation} 
By differentiating the left side of Equation \ref{appe:mgderiv}, I find
\begin{equation}\label{appe:mgderivused}
z\frac{\partial}{\partial z} \GMeijerG = \MeijerG{m}{n}{p}{q}{a_1-1,a_2, \dots, a_p}{b_1, \dots, b_q}{z} + (a_1-1)\GMeijerG \; .
\end{equation}
\subsection{The Radial Mass Profile}\label{app:radialmassprof} 
Here I will give the radial mass profile for a generic S\'{e}rsic profile following the treatment of \cite{MazureCapelato2002}. This is easily modified to give either the stellar or HI mass profiles using the correct $n$, $R_e$, and $\Sigma$. 

Define the S\'{e}rsic profile in terms of dimensionless quantities
\begin{eqnarray}\label{appe:dimsersic}
x &\equiv& \frac{r}{R_e}\\
s &\equiv& \frac{r}{R_e}\\
i(x) &=& \frac{I(R)}{I_e} = \exp[-b_n (x^{1/n} -1)] \\
\nu(s) &=& n(r) \frac{R_e}{I_e} = -\frac{1}{\pi}\int_s^{\infty} \frac{di}{dx} \frac{1}{\sqrt{x^2 - s^2}} dx \; .
\end{eqnarray}
Then, the deprojected radial luminosity profile is given by
\begin{equation} \label{appe:lumprofile}
L(s) = 4\pi \int_0^s s'^2 \nu(s') ds' \; .
\end{equation}
Since I am assuming that the mass follows the light, the radial mass profile is Equation \ref{appe:lumprofile} times the mass-to-light ratio, $\Upsilon$,
\begin{equation}
M(s) = 4\pi \Upsilon \int_0^s s'^2 \nu(s') ds' \; .
\end{equation}
\cite{MazureCapelato2002} find that the analytic solution to this integral for a S\'{e}rsic profile is
\begin{equation}\label{appe:ms}
M(s) = 2\pi \Upsilon c_1 s^{\frac{2n+1}{n}} \MeijerGFour{2n}{1}{1}{2n+1}{\{-\left(\frac{1}{2n}\right)\},}{\{\}}{\{\beta_s\},}{\{-\left(\frac{2n+1}{2n}\right)\}}{c_2s^2} \; ,
\end{equation}
where 
\begin{eqnarray}
c_1 &\equiv& \frac{b_n \exp[b_n]}{(2\pi)^n \sqrt{n}} \\
c_2 &\equiv& \left( \frac{b_n}{2n}\right)^{2n} \\
\beta_s &\equiv& \bigg\{ \left(\frac{j-1}{2n}\right)_{1\leq j\leq n} ; \left(\frac{j-2}{2n}\right)_{n+1\leq j \leq 2n}\bigg\} \; .
\end{eqnarray}

\subsection{EG Predictions}
To give values predicted by EG, I must find $\frac{dM}{dr}$. First, I differentiate Equation \ref{appe:ms} using Equation \ref{appe:mgderivused}, which gives
\begin{equation}
\frac{dM}{ds} = 4\pi \Upsilon c_1 s^{\frac{n+1}{n}} \MeijerGFour{2n}{0}{0}{2n}{\{\}}{\{\}}{\{\beta_s\}}{\{\}}{c_2s^2} \; .
\end{equation}
Now, I need to express $dM/ds$ and $M(s)$ in terms of $r$ instead. This is done using the definition for $s$ given in Equation \ref{appe:dimsersic} and accounting for the extra factor of $1/R_e$ from the change of variable in the derivative. However, I also need to account for how I began with a \textit{dimensionless} luminosity density by multiplying both $dM/ds$  and $M(s)$ by $I_eR_e^2$. Then,
\begin{eqnarray*}
M(r) &=& 2\pi c_1 \Sigma R_e^2 \left(\frac{r}{R_e}\right)^{\frac{2n+1}{n}} \MeijerGFour{2n}{1}{1}{2n+1}{\{-\left(\frac{1}{2n}\right)\},}{\{\}}{\{\beta_s\},}{\{-\left(\frac{2n+1}{2n}\right)\}}{c_2\left(\frac{r}{R_e}\right)^2} \; ,\\
\frac{dM}{dr} &=& 4\pi c_1 \Sigma R_e \left(\frac{r}{R_e}\right)^{\frac{n+1}{n}} \MeijerGFour{2n}{0}{0}{2n}{\{\}}{\{\}}{\{\beta_s\}}{\{\}}{c_2\left(\frac{r}{R_e}\right)^2} \; , \\
M_D^2(r) &=& \frac{\pi a_0 c_1 \Sigma R_e^2}{3G}\left(\frac{r}{R_e}\right)^{\frac{2n+1}{n}} r^2\left[\MeijerGFour{2n}{1}{1}{2n+1}{\{-\left(\frac{1}{2n}\right)\},}{\{\}}{\{\beta_s\},}{\{-\left(\frac{2n+1}{2n}\right)\}}{c_2 \left(\frac{r}{Re}\right)^2} + 2 \MeijerGFour{2n}{0}{0}{2n}{\{\}}{\{\}}{\{\beta_s\}}{\{\}}{c_2 \left(\frac{r}{Re}\right)^2} \right] \; ,
\end{eqnarray*}
where $\Sigma = I_e \Upsilon$.

\end{document}